\documentclass[twoside]{article}
\usepackage[accepted]{aistats2014}

\usepackage[sort&compress,numbers]{natbib}
\usepackage{comment}
\usepackage{hyperref}

\usepackage[ruled,lined]{algorithm2e}

\SetKwIF{If}{ElseIf}{Else}{if}{then}{else if}{else}{endif}

\usepackage{amsmath,bm,amsfonts,color,amsthm}
\usepackage{graphicx}

\DeclareMathOperator*{\argmax}{arg\,max}

\newcommand{\commentt}[1]{}

\newcommand{\lam}{\bm{\Lambda}}

\newtheoremstyle{remark}
{\topsep} 
{\topsep} 
{} 
{} 
{\bf} 
{:} 
{0.5em} 
{} 

\theoremstyle{remark}
\newtheorem{rem}{Remark}[section]

\begin{document}

%

%
\runningtitle{A Bayesian Approach to Graphical Record Linkage }

\twocolumn[


\aistatstitle{SMERED: A Bayesian Approach to Graphical \\ Record Linkage and De-duplication}

\aistatsauthor{ Rebecca C. Steorts \And Rob Hall \And Stephen E. Fienberg }

\aistatsaddress{ \texttt{ \{ beka, fienberg \} @cmu.edu}, Department of Statistics,
 Carnegie Mellon University, Pittsburgh, PA 15213 \\
  \texttt{ \{ rhall\} @etsy.com}, Etsy\\
 }]
 
%


\begin{abstract}
  We propose a novel unsupervised approach for linking records across
  arbitrarily many files, while simultaneously detecting duplicate records
  within files.  Our key innovation is to represent the pattern of links
  between records as a {\em bipartite} graph, in which records are directly
  linked to latent true individuals, and only indirectly linked to other
  records.  This flexible new representation of the linkage structure naturally
  allows us to estimate the attributes of the unique observable people in the
  population, calculate $k$-way posterior probabilities of matches across
  records, and propagate the uncertainty of record linkage into later analyses.
  Our linkage structure lends itself to an efficient, linear-time, hybrid Markov chain
  Monte Carlo algorithm, which overcomes many obstacles encountered by
  previously proposed methods of record linkage, despite the high dimensional
  parameter space. We assess our results on real and simulated data.
\end{abstract}
\section{Introduction}
\label{intro}
When data about individuals comes from multiple sources, it is essential to
match, or link, records from different files that correspond to the same
individual. Other names associated with record linkage are entity
disambiguation and coreference resolution, meaning that records which are
\emph{linked} or \emph{co-referent} can be thought of as corresponding to the
same underlying \emph{entity.}  Solving this 
problem is not just important as a preliminary to statistical analysis; the noise and distortions in typical data files make it a difficult, and intrinsically high-dimensional, 
problem \citep{Herzog_2007,lahiri_2005,Winkler_1999,winkler_2000}.
 
We propose a Bayesian approach to the record linkage problem based on a
parametric model that addresses matching $k$ files simultaneously and includes
duplicate records within lists. We represent the pattern of matches and non-matches as a bipartite graph, in which records are directly linked to the true but latent individuals which they represent, and only indirectly linked to other records.  Such {\em linkage structures} allow us to simultaneously solve three problems: record linkage, de-duplication, and estimation of unique observable population attributes.
The Bayesian paradigm naturally handles uncertainty about linkage, which poses a difficult challenge
to frequentist record linkage techniques.\footnote{\citet{liseo_2013} review Bayesian contributions to record linkage.}
Doing so permits valid statistical inference regarding posterior matching probabilities of records and propagation of errors as discussed in \S \ref{app}.

To estimate our model, we develop a hybrid MCMC algorithm, in the spirit of \cite{jain_2004}, which runs in linear time
in the number of records and the number of MCMC iterations, even in high-dimensional parameter spaces.
Our algorithm permits duplication across and within lists
but runs faster if there are known to be no duplicates within lists.
We achieve further
gains in speed using standard record linkage blocking techniques \citep{christen_2011}.

We apply our method to data from the National Long Term Care Survey (NLTCS), which tracked and surveyed approximately 20,000 people at five-year intervals.  At each wave of the survey, some individuals had died and were replaced by a new cohort, so the files contain overlapping but not identical sets of individuals, with no within-file duplicates.

\subsection{Related Work}
The classical 
work of \cite{fellegi_1969}
considered linking two files in terms of Neyman-Pearson hypothesis testing.  
Compared to this baseline, our approach is distinctive in that it handles multiple files, models distortion explicitly, offers a Bayesian treatment of uncertainty and error propagation, and employs a sophisticated graphical data structure for inference to latent individuals.  
Fellegi-Sunter methods based upon \cite{fellegi_1969} can extend to $k>2$ files \citep{sadinle_multi_1}, but they break down for even 
moderately large $k$
or complex data sets.  Moreover, they give
little information about uncertainty in matches, or about the true values of 
noise-distorted records.
The idea of modeling the distortion process
originates
with the ``Hit-Miss Model'' by \cite{copas_1990}, which
anticipates
  part of our model in \S \ref{model}. The specific distortion model we use is however closer to that introduced in 
  \cite{hall12}, as part of a nonparametric frequentist technique for matching $k=2$ files.  We differ from \cite{hall12} by introducing latent individuals and distortion through a Bayesian model.

Within the Bayesian paradigm,
most work has focused on specialized problems related to linking two files, which propagate uncertainty \cite{gutman_2013, liseo_2011, larsen_2001, belin_1995}.  These contributions, while valuable, do not easily generalize to multiple files and duplicate detection. 

Two recent papers \cite{domingos_2004, gutman_2013} are most relevant to the novelty of our work, namely the linkage structure.
To aid recovering information about the population from distorted records, \cite{gutman_2013} called for developing ``more sophisticated network data structures."  Our linkage graphs are such a data structure with the added benefit of permitting de-duplication and handling multiple files. Moreover, due to exact error propagation, our methods are also easily integrated with other analytic procedures. Algorithmically, the closest approach to our linkage structure 
is the graphical representation in \cite{domingos_2004}, for de-duplication within one file.  Their representation is a unaparatite graph, where records are linked to each other. Our use of a bipartite graph with latents individuals naturally fits in the Bayesian paradigm along with distortion. Our method is the first to handle
record linkage and de-duplication, while also modeling distortion and running in linear time. 
\section{Notation, Assumptions, and Linkage Structure}
\label{notation}
We begin by defining some notation,
where we have $k$ files or lists. For simplicity, we assume that all files
contain the same $p$ fields, which are all categorical, field $\ell$ having
$M_{\ell}$ levels.  We also assume that every record is complete.  (Handling
missing-at-random fields within records is a minor extension within the
Bayesian framework.)
Let $\bm{x}_{ij}$ be the data for the $j$th record in file $i$, where
$i=1,\ldots,k$,\; $j=1,\ldots,n_i$, and $n_i$ is the number of records in file
$i$; $\bm{x}_{ij}$ is a categorical vector of length $p$.  Let $\bm{y}_{j'}$
be the latent vector of
true field
values
for the $j'$th individual in the population (or rather aggregate sample), where
$j'=1,\ldots,N$, 
$N$ being
the total number of \emph{observed}
individuals from the population. 
$N$ could be as small as 1 if every
record 
in every file
refers to the same individual
or as large as
$N_{\max} \equiv \sum_{i=1}^k{n_i}$ if no datasets share any
individuals.

Now define the linkage structure
$\bm{\Lambda}=\{\lambda_{ij}\;;\;i=1,\ldots,k\;;\;j=1,\ldots,n_i\}$ where
$\lambda_{ij}$ is an integer from $1$ to $N_{\max}$ indicating which latent
individual the $j$th record in file $i$ refers to, i.e., $\bm{x}_{ij}$ is a
possibly-distorted measurement of $\bm{y}_{\lambda_{ij}}$.
Finally,
$z_{ij\ell}$ is $1$ or $0$ according to whether or not a particular
field $\ell$ is distorted in $\bm{x}_{ij}.$

As usual, we use $I$ for indicator functions (e.g., $I(x_{ij\ell}=m)$ is 1 when
the $\ell$th field in record $j$ in file $i$ has the value $m$), and $\delta_a$ for the distribution of a point mass at $a$ (e.g., $\delta_{y_{\lambda_{ij}\ell}}$).
The vector $\bm{\theta}_{\ell}$ of length $M_{\ell}$ denotes the multinomial
probabilities. 
%
For clarity, we always index as follows:
${i=1,\ldots,k;}$ ${j=1,\ldots,n_i;}$ ${j'=1,\ldots,N;}$ ${\ell=1,\ldots,p;}$ ${m=1, \ldots, M_{\ell}.}$
%


\subsection{Independent Fields Model}
\label{model}

 %
We assume that the files are conditionally independent, given the latent
individuals, and that fields are independent within individuals.
We formulate the following Bayesian parametric model, where the joint posterior is in closed form and we sample from the full conditionals
using a hybrid MCMC algorithm:

\begin{align*}
\bm{x}_{ij\ell}\mid\lambda_{ij},\bm{y}_{\lambda_{ij}\ell},z_{ij\ell},\bm{\theta}_\ell&\stackrel{\text{ind}}{\sim}
\begin{cases}
\delta_{\bm{y}_{\lambda_{ij}\ell}}&\text{ if }z_{ij\ell}=0\\
\text{MN}(1,\bm{\theta}_\ell)&\text{ if }z_{ij\ell}=1
\end{cases}\\
z_{ij\ell}&\stackrel{\text{ind}}{\sim}\text{Bernoulli}(\beta_\ell)\\
\bm{y}_{j'\ell}\mid\bm{\theta}_{j\ell}&\stackrel{\text{ind}}{\sim}\text{MN}(1,\bm{\theta}_\ell)\\
\bm{\theta}_\ell&\stackrel{\text{ind}}{\sim}\text{Dirichlet}(\bm{\mu}_\ell)\\
\beta_\ell&\stackrel{\text{ind}}{\sim}\text{Beta}(a_\ell,b_\ell) \\
\pi(\lam) &\propto 1,
\end{align*}
where
$a_\ell,b_\ell,$ and $\bm{\mu}_\ell$ are all known, and MN denotes the Multinomial distribution.

\begin{rem}
  We assume that every legal configuration of the $\lambda_{ij}$ is equally
  likely a~priori.
  This implies a
  non-uniform prior on
  related quantities,
  such as the number of individuals in the data.  
  The uniform
  prior on $\bm{\Lambda}$ is convenient, since constructing either a subjective
  or an alternative objective prior is unclear.
 \textcolor{black}{
A uniform distribution on one quantity, i.e. $\Lambda$, implies a non-uniform distribution on other, related quantities (such as $N$).  Making every entry in the matrix $\lambda_{ij}$ uniformly distributed on $1, 2, \ldots N_{\max}$ implies that the distribution of $N$, a function of $\Lambda$, is not uniform on $1, 2, \ldots N_{\max}$. This is a long-standing problem with ``non-informative priors" \cite{kass_1996}.}
\end{rem}

Deriving the joint posterior and conditional distributions is now mostly
straightforward.  One subtlety, however, is that $\bm{y}$, $\bm{z}$ and
$\bm{\Lambda}$ are all related, since if $z_{ij\ell}=0$, then it must be the
case that $y_{\lambda_{ij}\ell}=x_{ij\ell}$.
%
Taking this into account,
the joint posterior is
\begin{align*}
\label{joint}
&\pi(\bm{\Lambda},\bm{y},\bm{z},\bm{\theta},\bm{\beta}\mid\bm{x})\\
&\propto
\prod_{i,j,\ell, m}
\left[(1-z_{ij\ell})\delta_{y_{\lambda_{ij}\ell}}(x_{ij\ell})
\;\;+\;\;
z_{ij\ell}\theta_{\ell m}^{I(x_{ij\ell} = m)}\right]\notag\\
&\phantom{\propto{}}\times
\prod_{\ell, m}
\theta_{\ell m}^{\mu_{\ell m}+\sum_{j'=1}^N I(y_{j'l}=m)}\notag\\
&\phantom{\propto{}}\times
\prod_{\ell}
\beta_\ell^{a_\ell-1+\sum_{i=1}^k\sum_{j=1}^{n_i}z_{ij\ell}} \notag \\
&\times
(1-\beta_\ell)^{b_\ell-1+\sum_{i=1}^k\sum_{j=1}^{n_i}(1-z_{ij\ell})}. \notag
\end{align*}
We suppress
derivation of the full conditionals, but note that the full conditionals of 
$\bm{y}$,
$\bm{z}$ and $\bm{\Lambda}$ always obey their logical dependence, and therefore never condition on impossible events.
The full conditional of $\bm{\Lambda}$ must reflect whether or not
there are duplicates within files.  If we define $R_{ij^{\prime}} = \left\{ j : \lambda_{ij} = j^{\prime}\right\},$ then not having within-file duplicates means that $R_{ij^\prime}$ must be either $\emptyset$ or a single record, for each $i$ and $j^{\prime}$. Graphically, this means
allowing
or forbidding links from a latent individual to multiple records within one
file. 

\subsection{Split and MErge REcord linkage and De-duplication (SMERED)
Algorithm}
\label{alg}

Our main goal is estimating the posterior distribution of the linkage (i.e.,
the clustering of records into individuals). The simplest way of accomplishing
this is via Gibbs sampling.  We could iterate through the records, and for each
record, sample a new assignment to an individual (from among the individuals
represented in the remaining records, plus an individual comprising only that
record).  However, this requires the quadratic-time checking of proposed
linkages for every record.
%
Thus, instead of  Gibbs sampling, we use a hybrid MCMC algorithm to explore the space of possible linkage structures, which allows our algorithm to run in linear time.

Our hybrid MCMC
takes
advantage of split-merge moves, as done in \cite{jain_2004},
which avoids the problems associated with
Gibbs sampling,
even though the number of parameters grows with the number of records.
This is accomplished via proposals that can traverse the state space quickly
and frequently visit high-probability modes, since the algorithm splits or
merges records in each update, and hence, frequent updates of the Gibbs sampler
are not necessary.

Furthermore, a common technique in record linkage is to require an exact match
in certain fields (e.g., birth year) if records are to be
linked.
This technique
of \emph{blocking}
can greatly reduce the number of possible 
links
between records  (see e.g., \cite{winkler_2000}).
Since blocking gives up on finding truly co-referent records which disagree
on those fields, it is best to block on
fields that have little or no distortion.
We block on the fairly reliable fields of sex and
birth year in our application to the NLTCS below.
A strength of
our model
is that it 
incorporates blocking organically.
Setting
$b_\ell=\infty$ for a particular field $\ell$
forces
the distortion probability for 
that field to zero.
This requires 
matching
records to agree on the $\ell$th field, 
just like blocking.

We now discuss how the split-merge process links records to records, which it
does by
assigning
records to latent individuals.  Instead of sampling 
assignments
at the record level, we do so at the individual level.  Initially, each record
is assigned to a unique individual.  On each iteration,
we
choose two records at random.  If 
the pair
belong to 
\emph{distinct}
latent individuals, then we propose merging those individuals to form a single
new latent individual (i.e., we propose that those records are co-referent).
On the other hand, if the two records belong to the \emph{same} latent
individual, then we propose splitting it into two new latent individuals, each
seeded with one of the two chosen records, and the other records randomly
divided between the two.  Proposed splits and merges are accepted based on the
Metropolis-Hastings ratio and rejected otherwise.

To choose the pair of records,
one option is to sample uniformly from among all possible pairs.  However, this
is not ideal, for two reasons.  First,
most 
pairs of records are extremely unlikely to 
match
since they agree on few, if any, fields.  Frequent proposals to merge such
records are wasteful. Therefore, we employ blocking,
and only consider pairs of records within the same block.
Second,
sampling
from all possible pairs of records will sometimes lead to proposals to merge
records in the same list.  If we permit duplication within lists, then this is
not a problem.  However, if we know (or assume) there are no duplicates within
lists, we should avoid wasting time on such pairs.
The no-duplication version of our algorithm does precisely this. (See Algorithm
\ref{alg:smered} for pseudocode.) \textcolor{black}{When there are no duplicates within files, we call the SMERE (Split and MErge REcord linkage) algorithm, which enforces the restriction that $R_{ij^{\prime}}$ must be either $\emptyset$ or a single record.  This is done through limiting the proposal of record pairs to those in distinct files; the algorithm otherwise matches SMERED.}
  
   \begin{algorithm}[t!]
   \DontPrintSemicolon
     \BlankLine
   \KwData{$\bm{X}$ and hyperparameters}
    Initialize the unknown parameters $\bm{\theta}, \bm{\beta}, \bm{y}, \bm{z},$ and $\bm{\Lambda}.$
   \BlankLine
   \For{$i \leftarrow 1$ \KwTo $S_G$} {
    	\For{$j \leftarrow 1$ \KwTo $S_M$} {\
		\For{$t \leftarrow 1$ \KwTo $S_T$} {\
			Draw records $R_1$ and $R_2$ uniformly and independently at random. \\
			\If{$R_1$ and $R_2$ refer to the same individual}{propose splitting that individual, shifting $\bm{\Lambda}$ to $\bm{\Lambda^{\prime}}$\\
			 }
			\Else{propose merging the individuals $R_1$ and $R_2$ refer to, shifting $\bm{\Lambda}$ to $\bm{\Lambda^{\prime}}$\\
			}
			$r \leftarrow \min{
			\left\{
			 1, \frac{\pi(\bm{\Lambda}^{\prime},\bm{y},\bm{z},\bm{\theta},\bm{\beta}|\bm{x}) }{ \pi(\bm{\Lambda},\bm{y},\bm{z},\bm{\theta},\bm{\beta}|\bm{x})} \right\} 
			 } $\\
			Resample $\bm{\Lambda}$ by accepting proposal with Metropolis probability $r$ or rejecting with probability $1-r.$
			}
	Resample  $\bm{y}$ and $\bm{z}.$
	
	}
	Resample $\bm{\theta}, \bm{\beta}.$
   } 
  \BlankLine
   \Return{$\bm{\theta}|\bm{X}, \bm{\beta}\bm{X}, 
   \bm{y}|\bm{X}, \bm{z}|\bm{X},$ and $\bm{\Lambda}|\bm{X}.$}
   \caption{Split and MErge REcord linkage and De-duplication (SMERED)}
   \label{alg:smered}
 \end{algorithm}
 

\subsubsection{Time Complexity}
\label{time}

Scalability is crucial to any record linkage algorithm 
Current approaches typically run in polynomial (but super-linear) time in $N_{\max}$. (The method of \cite{sadinle_multi_1} is $O(N_{\max}^k)$, while that of  \cite{domingos_2004} finds the maximum flow in an $N_{\max}$-node graph, which is $O(N_{\max}^3)$, but independent of $k$.)
In contrast, our algorithm is linear in both $N_{\max}$ and MCMC iterations.

Our
running time is proportional to the number of Gibbs iterations $S_G,$ 
so
we focus on the time
taken by one
Gibbs step. 
Recall the notation from \S \ref{notation}, and define $M =
\frac{1}{p} \sum_{\ell=1}^p M_{\ell}$ as the average number of possible values
per field ($M \geq 1$).
The time taken by a Gibbs step
is dominated by
sampling from the
conditional distributions. Specifically, sampling
$\bm{\beta}$ and $\bm{y}$ are both $O(p N_{\max})$; sampling $\bm{\theta}$ is $O(pMN) + O(pN_{\max}) = O(pMN)$, as is sampling $\bm{z}$.  Sampling $\bm{\Lambda}$ is $O(pN_{\max}M)$ if done carefully. Thus, all these samples can be drawn in 
time linear in $N_{\max}$. 


Since there are $S_M$ Metropolis steps within each Gibbs step and each Metropolis step updates $\bm{y}$, $\bm{z}$, and $\bm{\Lambda}$, the time needed for the Metropolis part of one Gibbs step is $O(S_MpN_{\max}) + O(S_MpMN) + O(S_MpN_{\max}M).$ Since $N \leq N_{\max},$  the run time becomes 
$O(pS_M N_{\max}) + O(MpS_M N_{\max}) = O(MpS_M  N_{\max}). $ On the other hand, the updates for $\bm{\theta}$ and $\bm{\beta}$ occur once each Gibbs step implying the run time is $O(pMN) + O(pN_{\max}).$ Since $N \leq N_{\max},$ the run time becomes $O(pM  N_{\max} + p  N_{\max})  = O(p  M N_{\max} ).$ The overall run time of a Gibbs step is 
$O(pM  N_{\max} S_M) +  O(p  M N_{\max} ) = O(pM  N_{\max} S_M).$ Furthermore, for $S_G$ iterations of the Gibbs sampler, the algorithm is order
$O(pM  N_{\max} S_G S_M).$ If $p$ and $M$ are all much
less than $N_{\max}$, we find that the runtime is $O( N_{\max} S_G S_M).$


\textcolor{black}{Another important consideration is the number of MCMC steps needed to produce Gibbs samples that form an adequate approximation of the true posterior.  This issue depends on the convergence properties \textcolor{black}{(actual rate of convergence)} of the hybrid Markov chain used by the algorithm, which are beyond the scope of the present work.} \textcolor{black}{Convergence diagnostics  for our application to the NLTCS and hyperparameter sensitivity is discussed in Appendix \ref{app:conv}}.

\subsection{Posterior Matching Sets and Linkage Probabilities}
%
%
In a Bayesian framework, the output of record linkage is not a deterministic set of matches between records, but a probabilistic description of how likely records are to be co-referent, based on the observed data.
Since we are linking multiple files at once, we propose a range of
posterior matching probabilities:
the posterior probability of linkage
between
two arbitrary records and more generally
among
$k$ records, the posterior probability given a set of records that they are
linked, and the posterior probability that a given set of records is a maximal
matching set (which will be defined later).

Two records $(i_1,j_1)$ and $(i_2,j_2)$ {\em match} if they
point to the same latent individual, so $\lambda_{i_1j_1} = \lambda_{i_2j_2}.$
The posterior probability of a match can be computed from the $S_G$ MCMC
samples:
$$
P(\lambda_{i_1j_1} = \lambda_{i_2j_2} | \bm{X}) = \frac{1}{S_G}\sum_{h=1}^{S_G}
I(\lambda_{i_1j_1}^{(h)} = \lambda_{i_1j_2}^{(h)}).$$ 
A one-way match
is
when an individual appears in only one of the $k$ files, while
a two-way match
is
when an individual appears in exactly two of the $k$ files, and so on (up to $k$-way
matches).  We approximate the posterior probability of arbitrary one-way,
two-way, \ldots, $k$-way matches as the
ratio of
the number of times those matches happened in the posterior sample
to $S_G$.

\textcolor{black}{Although probabilistic results and interpretations provided by the Bayesian paradigm are useful both quantitatively and conceptually, 
we often
 report
a 
point estimate of the linkage structure.  
 Thus, we face the question of how to condense the overall posterior distribution of $\bm\Lambda$ into a single estimated linkage structure.}

\textcolor{black}{Perhaps the most obvious approach is to set some threshold~$v$, where $0<v<1$, and to declare (i.e., estimate) that two records match if and only if their posterior matching probability exceeds~$v$.  This strategy is useful if only a few specific pairs of records are of interest, but its flaws are exposed when we consider the coherence of the overall estimated linkage structure implied by such a thresholding strategy.  Note that the true linkage structure is \emph{transitive} in the following sense: if records A and B are the same individual, and records B and C are the same individual, then records A and C must be the same individual as well.  However, this requirement of transitivity is in no way enforced by the simple thresholding strategy described above.  Thus, a more sophisticated approach is required if the goal is to produce an estimated linkage structure that preserves transitivity.}

\textcolor{black}{To this end, it is useful to define a new concept.}
A set of records $\mathcal{A}$
is
a \emph{maximal matching} set (MMS) if
\textcolor{black}{every record in the set has}
the same value of ${\lambda}_{ij}$ and
\textcolor{black}{no record outside the set has that}
value of ${\lambda}_{ij}.$ Define $\bm{\Omega}(\mathcal{A}, \bm{\Lambda})
:=\bm{\Omega}_{\mathcal{A}, \bm{\Lambda}} $ to be
1 if 
$\mathcal{A}$ is
an MMS in $\bm{\Lambda}$ and 0 otherwise:
$$\bm{\Omega}_{\mathcal{A}, \bm{\Lambda}} = \sum_{j^{\prime}}{\left( \prod_{(i,j) \in \mathcal{A}}{I(\lambda_{i j}=j^{\prime})}\prod_{(i,j)\not\in\mathcal{A}}{I(\lambda_{ij}\neq j^{\prime}})\right)}.$$
Essentially, the MMS contains all the records which match some
particular latent individual, though which individual
is irrelevant.
%
Given a set
of records~$\mathcal{A}$, the posterior probability that it is an MMS in $\bm\Lambda$ is simply
%
\begin{align*}
P(\bm{\Omega}_{\mathcal{A}, \bm{\Lambda}} =1)
&= 
\frac{1}{S_G}\sum_{h=1}^{S_G}
{\bm{\Omega}(\mathcal{A},\bm{\Lambda}^{(h)})}.
\end{align*}


\textcolor{black}{The MMSs allow a sophisticated method of preserving transitivity when estimating a single overall linkage structure.
For any record $(i,j)$, its \emph{most probable MMS}~$\mathcal{M}_{ij}$ is the set containing $(i,j)$ with the highest posterior probability of being an MMS, i.e.,
$$
\mathcal{M}_{ij}:=
\argmax_{\mathcal{A}:(i,j)\in\mathcal{A}}
P(\bm{\Omega}_{\mathcal{A}, \bm{\Lambda}} =1).
$$
Next, a \emph{shared most probable MMS} is a set that is the most probable MMS of all records it contains, i.e., a set $\mathcal{A}^\star$ such that $\mathcal{M}_{ij}=\mathcal{A}^\star$ for all $(i,j)\in\mathcal{A}^\star$.  We then estimate the overall linkage structure by linking records if and only if they are in the same shared most probable MMS.  The resulting estimated linkage structure is guaranteed to have the transitivity property since (by construction) each record is an element of at most one shared most probable MMS.}


\subsection{Functions of Linkage Structure }
\label{sub:function}


\textcolor{black}{The output of the Gibbs sampler also allows us to estimate the value of any function of the variables, parameters, and linkage structure by computing the average value of the function over the posterior samples.  For example, estimated summary statistics about the population of latent individuals are straightforward to calculate.  Indeed, the ease with which such estimates can be obtained is yet another benefit of the Bayesian paradigm, and of MCMC in particular.}

\section{Assessing Accuracy of Matching and Application to NLTCS}
\label{app}

We test
  our model on data from the NLTCS, a
  longitudinal study of the health of elderly (65+)  individuals (\url{http://www.nltcs.aas.duke.edu/}).
  The NLTCS was conducted approximately every six years, with each
  wave containing roughly 20,000 individuals.  Two aspects of the NLTCS make it
  suitable for our purposes: individuals were tracked from wave to wave with
  unique identifiers, but at each wave, 
  many
  patients had
  died (or otherwise left the study) and were replaced by newly-eligible
  patients.  We can test the ability of our model to link records across
  files by seeing how well it is able to track individuals across waves, and
  compare its estimates to the ground truth provided by the unique
  identifiers.

To show how little information our method needs to find links across files,
  we gave it access to only four variables, all known to be noisy: full date of birth, sex, state of
  residence, and the regional office at which the subject was interviewed. 
  \textcolor{black}{We treat all fields as categorical.}
  We linked individuals across the 1982, 1989 and 1994 survey
  waves.\footnote{The 
  other three waves
  used different questionnaires and  are not
  strictly comparable.} 
  Our model had little information on which to link, and not \emph{all} of its
   assumptions strictly hold
 (e.g., individuals can move between states across waves).
  %
  %
  We demonstrate our method's validity using error rates, confusion matrices, 
  posterior matching sets and linkage probabilities, and estimation of the unknown 
  number of observed individuals from the population.
  %

Appendix \ref{sec:sim} provides a simulation study of the NLTCS with varying levels of distortion at the field level.  We conclude from this that SMERE is able to handle low to moderate levels of distortion (Figure \ref{distort_six}).  Furthermore, as distortion increases, so do the false negative rate (FNR) and false positive rate (FPR) (Figure \ref{distort}).
  
\subsection{Error Rates and Confusion Matrix}
\label{sub:basic}
Since we have unique identifiers for the NLTCS, we can see how
accurately our model matches records.  A \emph{true link} is a match between
records which really do refer to the same latent individual; a \emph{false link} is a
match between records which refer to different latent individuals; and a \emph{missing
  link} is a match which is not found by the model.  Table~\ref{truelinks_dup}
gives posterior means for the number of true, false, and missing links.
%
For the NLTCS, 
the FNR is $0.11,$ 
while the FPR is $0.046,$ when we block by date of birth
year (DOB) and sex.

\begin{figure}[htbp]
\begin{center}
\includegraphics[width=0.8\columnwidth]{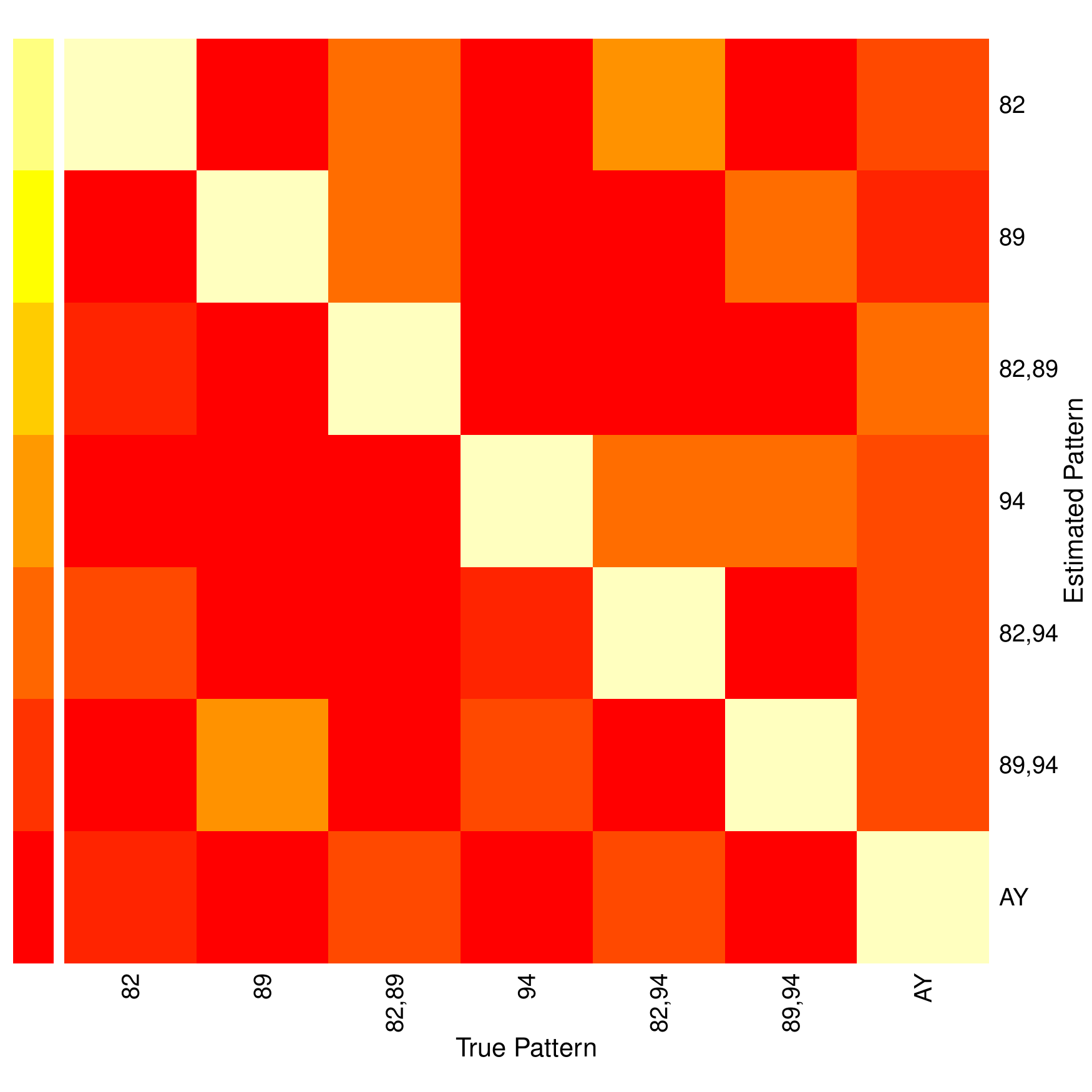}
\caption{Heatmap of relative probabilities from the confusion matrix, running
  from yellow (most probable) to dark red (probability 0).  The largest
  probabilities are on the diagonal, showing that the linkage patterns
  estimated for records are correct with high probability.  Mis-classification
  rates are low and show a tendency to under-link rather than over-link.}
\label{heatmap}
\end{center}
\end{figure}

More refined information about linkage errors comes from a confusion matrix,
which compares records' estimated and actual linkage patterns
(Figure \ref{heatmap} and  Appendix
\ref{app:confusion}, Table~\ref{confusion}).
%
Every row in the confusion matrix is diagonally dominated, indicating that
correct classifications are overwhelmingly probable.  The largest off-diagonal
entry, indicating a mis-classification, is $0.07$.  For instance, if a record
is estimated to be in both the 1982 and 1989 waves, it is 90\% probable that
this estimate is correct.  If the estimate is wrong, the truth is most probably
that the record is in all waves (4.4\%), followed by the 1982 wave alone
(1.4\%) and waves 1982 and 1994 (0.15\%), and then other patterns with still
smaller probability.
%
\subsection{Example of Posterior Matching Probabilities}
We wish to search for sets of records that match record 10084 in 1982. In the
posterior samples of $\bm{\Lambda}$, this record is part of three maximal
matching sets that occur with nonzero estimated posterior probability, one with high and two with low posterior matching probabilities
(Table \ref{postmatch}).  This record has a posterior probability of $0.995$ of
simultaneously matching both record 6131 in 1989 and record 5583 in 1994.  All
three records denote a male, born 07/01/1910, visiting office 25 and residing
in state 14.  The unique identifiers show that these three records are in fact
the same individual.
  %
If we threshold 
matching sets, ignoring ones of low posterior probability,
we would simply return the set of records in last column of
Table~\ref{postmatch}.

\subsection{Estimation of Attributes of Observed Individuals from the Population}
\label{sub:est-function}

\begin{figure}[h]
\begin{center}
\includegraphics[scale = 0.4]{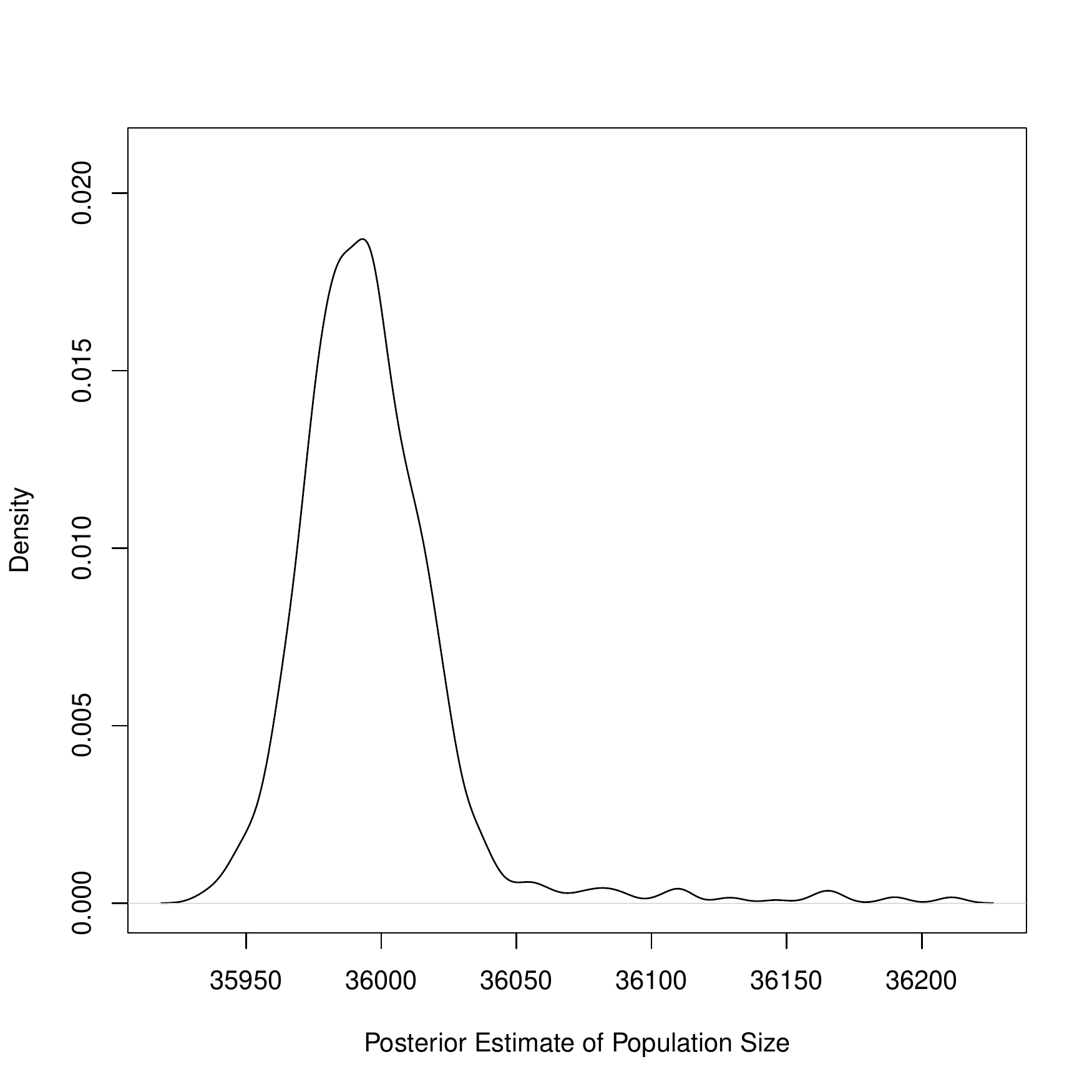}
\caption{Posterior density of the number of observed unique individuals $N.$}
\label{n}
\end{center}
\end{figure}

The number of observed unique individuals $N$ is easily inferred from the posterior of
$\bm{\Lambda}|\bm{X},$ since $N$ is simply the number of unique values in
$\bm{\Lambda}.$
Defining $N|\bm{X}$ to be the posterior distribution of~$N,$ we
can find this by applying a function to the posterior distribution on $\bm{\Lambda}$, 
as discussed in \S \ref{sub:function}.
(Specifically, $N=|\#\bm{\Lambda}|$, where $\#\bm{\Lambda}$ maps $\bm{\Lambda}$ to its set of unique entries, and $|A|$ is the cardinality of the set $A$.)
Doing so, the posterior distribution of $N|\bm{X}$ is given
in Figure (\ref{n}).  Also, $\hat{N}$:= $E(N|\bm{X}) = 35,992$ with a
posterior standard error of 19.08. Since the true number of observed unique individuals is 34,945, we are overmatching, which leads to an overestimate of~$N$. This phenomenon most likely occurs due to
patients migrating between states across the three different waves. It is
difficult to improve this estimate since we do not have additional information
as just described above.  

We can also estimate attributes of sub-groups.
  For example, we can
estimate the number of individuals within each wave or combination of waves---that is, the number of individuals with any given linkage pattern.
(We summarize these estimates here with posterior expectations alone, but the
full posterior distributions
are easily computed.)
For example, the posterior expectation for the number of individuals appearing in
lists $i_i$ and $i_2$ but not $i_3$ is approximately
$$
\frac{1}{S_G} \sum_{h=1}^{S_G} \sum_{j'} 
I\left( \left | R_{i_1j'}^{(h)} \right | = 1 \right)
I\left ( \left | R_{i_2j'}^{(h)} \right | = 1 \right)
I\left( \left | R_{i_3j'}^{(h)} \right | = 0\right ).
$$
(Note that the inner sum is a function of $\bm{\Lambda}^{(h)}$, but a very complicated one to express without the $R_{ij}$.)

Table~\ref{de-smered_waves} reports the 
posterior means for the
overlapping waves and each single wave of the NLTCS and compares this to the
ground truth.
In the first
wave
(1982), our estimates perform
exceedingly well with relative error of 0.11\%, however, as waves cross and we
try to match people based on limited information,
the relative errors range from 8\% to 15\%. This is not surprising, since as
patients age, we expect their proxies to respond, making patient data more
prone to errors.  Also, older patients may move across states, creating further
matching dilemmas.
We are unaware of any alternative algorithm that does better on this data with
only these fields available.  
Given 
these results, 
and considering how little field information we allowed it to use for matching,
we find that our
model performs overall very well.

%
%

\section{De-duplication}
\label{dedup}

Our application of SMERE to the NLTCS 
assumes
that each list
had no
 duplicates, however, many other applications will contain duplicates within lists.
We showed in \S \ref{model} that we can theoretically handle de-duplication across and within lists.  We apply SMERE with  de-duplication (SMERED) to the NLTCS by (i) running SMERED on the three waves to show that 
the algorithm does not falsely detect duplicates when there really are none, and (ii) combining all the lists into one file, hence creating many duplicates, to show that SMERED can find them. 



\subsection{Application for NLTCS}

We combine the three files of the NLTCS mentioned in \S \ref{app} which contain 22,132 duplicate records out of 57,077 total records.  We run SMERED on settings (i) and (ii),
evaluating accuracy with the unique IDs.

In the the case of running SMERED on the three waves, we compare our results
of SMERED and SMERE to that under ground truth (Table
\ref{de-smered_waves}).  In the case of the NLTCS, compiling all
three files together and running the three waves separately under SMERED
yields similar results, since we match on similar covariate
information. There is no covariate information to add to from thorough
investigation to improve our results, except under simulation
study. Specifically, when running SMERED for the three files, the FNR is
0.11 and is 0.38 for FPR, while its FNR and FPR is 0.11 AND 0.37 for the one
compiled file. We contrast this with the FNR of 0.11 and FPR of 0.046 under
SMERE for the three waves (Table \ref{truelinks_dup}).

The dramatic increase in the FPR and number of false links shown in Table
\ref{de-smered_waves} is explained by how few field variables we match on.
Their small number means that there are many records for different individuals
that have identical or near-identical values. On examination, there are $2,558$
possible matches among ``twins,'' records which agree exactly on all attributes
but have different unique IDs.  Moreover, there are 353,536 ``near-twins,'' 
pairs of records that have different unique IDs but match on all but
one attribute.  This illustrates why the matching problem is so hard for the
NLTCS and other data sources like it, where survey-responder information like
name and address are lacking.  However, if it is known that each file contains
no duplicates, there is no need to consider most of these twins and near-twins
as possible matches.

\section{Discussion}
\label{disc}

We have made two contributions in this paper.  The first 
is to frame record linkage and de-duplication simultaneously, namely
linking observed records to latent individuals and representing the linkage
structure via $\bm{\Lambda}$.  The second contribution is our specific
parametric Bayesian model, which, combined with the linkage structure, allows
for efficient inference and exact error rate calculation.  Moreover, this
allows for 
easy
integration with capture-recapture methods, where error propogation is
\emph{exact.}
As with any parametric model, its assumptions only apply to certain problems,
but it also
serves as a starting point for more elaborate models, e.g., 
with
missing fields, data fusion, complicated string fields, population
heterogeneity, or dependence across fields, across time, or across individuals.
Within the Bayesian paradigm, 
such model expansions
will lead to
larger
parameter spaces, and therefore call for computational speed-ups, perhaps via online learning, 
variational inference, or approximate Bayesian computation.

Our work serves as a first basis for
solving
record linkage problems using a noisy Bayesian model, a linkage structure that
can handle large-scale databases, and a model that simultaneously combines
record linkage and de-duplication for
arbitrarily many
files. We
hope
that our 
approach
will encourage the emergence of new record linkage approaches, \textcolor{black}{extensions of our method to non-categorical fields, and applications}
along with more state-of-the-art algorithms for this kind of high-dimensional
data.


\paragraph*{Acknowledgements}This research was supported by NSF Census Research Network (NCRN), Research Training Grant (NSF),  Singapore National Research Foundation (NRF) under its International Research Centre @ Singapore Funding Initiative and the Interactive Digital Media Programme Office (IDMPO) to the Living Analytics Research Centre (LARC).
We thank the referees, the NCRN research node at CMU, Chris Genovese, Cosma Shalizi, Doug Sparks for providing helpful comments.

\begin{table}[h]
\begin{center}
\begin{tabular}{c|ccc}
sets of records & 
 1.10084       &        3.5583; 1.10084  &    3.5583; 1.10084; 2.6131  \\ \hline
posterior probability & 0.001 & 0.004  & 0.995\\
\end{tabular}
\end{center}
\caption{Example of posterior matching probabilities for record 10084 in 1982}
\label{postmatch}
\end{table}

\begin{table}[h]
\begin{center}
\hspace*{-2em}
\begin{tabular}{c|ccc|ccc|c}
 & 82& 89 &94& 82, 89 &89, 94& 82, 94  & 82, 89, 94 \\ \hline
NLTCS (ground truth)  &  7955  & 2959 & 7572
& 4464  
& 3929  
&  1511  
&  6114 \\ 
Bayes  Estimates$_{\text{SMERE}}$ & 7964  & 3434.1 & 8937.8
& 4116.9
& 4502.1  
&   1632.2 
&  5413\\
Bayes  Estimates$_{\text{SMERED}}$
 &7394.7  & 3009.9 & 6850.4
& 4247.5
& 3902.7  
&   1478.7 
&  5191.2\\
Relative Errors$_{\text{SMERE}}$  (\%)  & 
 0.11  &16.06 & 18.04 & $-$7.78 & 14.59 &  8.02 & $-$11.47\\
 Relative Errors$_{\text{SMERED}}$  $(\%)$  & 
$-$7.04  & 1.72 &  $-$9.53 & $-$4.85 & $-$0.67 & $-$2.14 & $-$15.09

\end{tabular}
\end{center}
\caption{Comparing NLTCS (ground truth) to the Bayes estimates of matches for SMERE and SMERED}
\label{de-smered_waves}
\end{table}

\begin{table}
\begin{center}
\begin{tabular}{c|ccc|cc}
& False links & True Links & Missing Links  & FNR & FPR  \\ \hline
NLTCS (ground truth) & 0 & 28246 & 0 & 0 & 0 \\
Bayes Estimates$_{\text{SMERE}}$ & 1298.9 & 25196 & 3050 & 0.11 & 0.05\\
Bayes Estimates$_{\text{SMERED}}$ &  10595 & 24900 & 3346 &  0.09 & 0.37\\
\end{tabular}
\end{center}
\caption{False, True, and Missing Links for NLTCS under blocking sex and DOB year where the Bayes estimates are calculated in the absence of duplicates per file and when duplicates are present (when combining all three waves). Also, reported FNR and FPR for NLTCS, Bayes estimates.}
\label{truelinks_dup}
\end{table}%



\twocolumn

\bibliographystyle{ims}
\bibliography{chomp}

\clearpage
\onecolumn
\appendix

\section{Simulation Study}
\label{sec:sim}  
\begin{figure}[h]
\begin{minipage}[t]{0.4\linewidth}
\centering
\includegraphics[width=\textwidth]{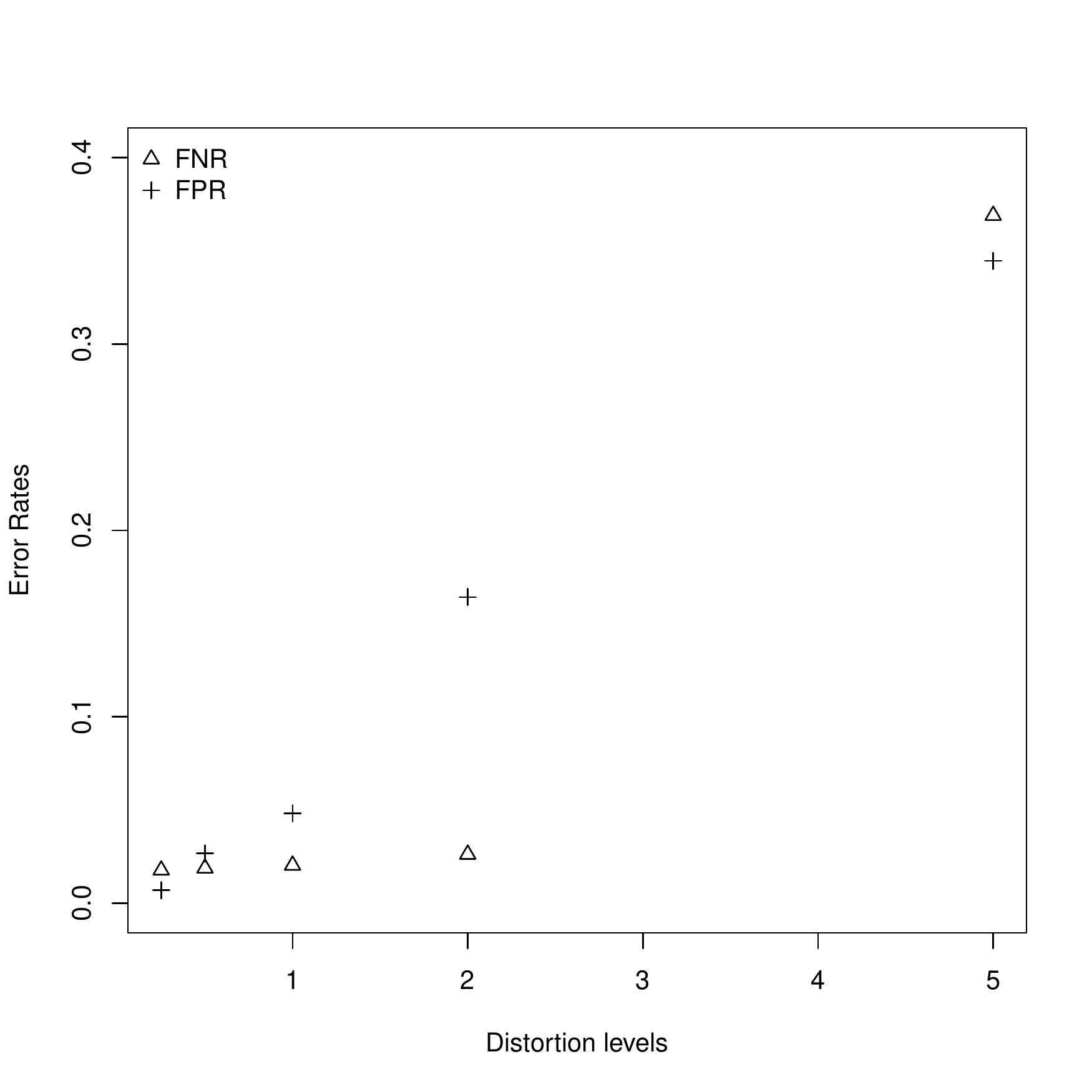}
\caption{FNR and FPR plotted against 5 levels of distortion, where the former (plusses) shows near linear relationship and latter shows exponential one (triangles).}
\label{distort}
\end{minipage}
\hspace{0.5cm}
\begin{minipage}[t]{0.4\linewidth}
\centering
\includegraphics[width=\textwidth]{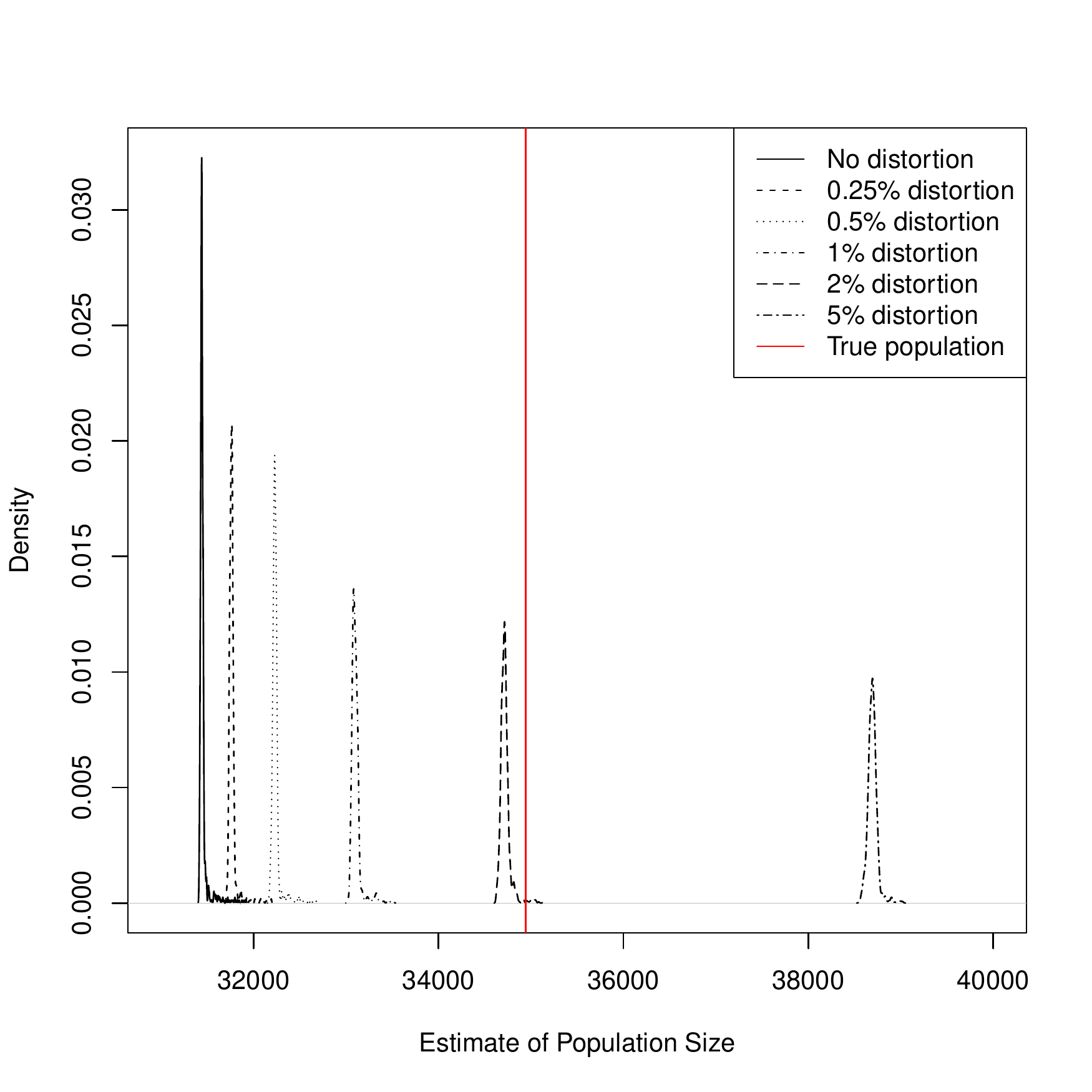}
\caption{Posterior density estimates for 6 levels of distortion (none, 0.25\%, 0.5\%, 1\%, 2\%, and 5\%) compared to ground truth (in red). As distortion increases (and approaches 2\% per field), we undermatch $N$, however as distortion quickly increases to high levels (5\% per field), the model overmatches. This behavior is expected to increase for higher levels of distortion. The simulated data illustrates that under our model, we are able to capture the idea of moderate distortion (per field) extremely well.}
\label{distort_six}
\end{minipage}
\end{figure}

We provide a simulation study based on the model in \S \ref{model} and we
simulate data from the NLTCS based on our model, with varying levels of
distortion. The varying levels of distortion (0, 0.25\%, 0.5\%, 1\%, 2\%, 5\%)
associated with the simulated data are then run using our MCMC algorithm to
assess how well we can match under ``noisy data.''  Figure \ref{distort}
illustrates an approximate linear relationship with FNR and the distortion
level,
while we see an near-exponential relationship
between FPR and the distortion level.
%
Figure~\ref{distort_six} demonstrates that for moderate distortion levels (per
field), we can estimate the true number of observed individuals 
extremely well via estimated
posterior densities. However, once the distortion is too
\emph{noisy}, our model has trouble recovering  this value.

In summary, 
as records become more noisy or distorted, our matching algorithm
\textcolor{black}{typically matches less than 80\% of the individuals}.  Furthermore, once the
distortion is around 5\%, we can only hope to recover approximately 65\%
of the individuals. 
Nevertheless, this degree of accuracy is in fact quite encouraging given the noise inherent in the data and given the relative lack of identifying variables on which to base the matching.

\section{Convergence Diagnostics and Hyperparameter Sensitivity} 
\label{app:conv}

\textcolor{black}{As for convergence diagnostics, for $S_G,$ our standard for NLTCS when running SMERE was to set $S_G = S_M = 10^5$,  after fixing on a burn-in of 1000 steps and a thinning the chain by 100 iterations from pilot runs. For SMERED, we used  $S_G = 10^5$ and $S_M = 10000.$ Moreover, our simulation study (Appendix A) varies $a_{\ell}$ and $b_{\ell}$ but we do not varying $\mu_l$ away from a uniform; if users have a priori knowledge regarding some idea about the expected distribution of categories, though, this could be incorporated fairly directly. For the NLTCS study itself, we set the parameters of $\beta$ are $a_{\ell} = 5 $ and $b_{\ell} = 10$ and took $\mu_{\ell} = 1,$ corresponding to equivalent to a uniform distribution over the $M_{\ell}-1$ simplex.  }

\clearpage

\section{Confusion Matrix for NLTCS}
\label{app:confusion}

\begin{table}[htdp]
\begin{center}
\hspace*{-2em}
\small
\begin{tabular}{c|ccc|ccc|c|c}
Est vs Truth & 82& 89 &82,89& 94 &82, 94 & 89, 94  & AY & RS\\ \hline
82 & 8051.9& 0.0& 385.1 &0.0 &162.9 & 0.0& 338.6  & 8938.5\\

89 & 0.0 & 2768.4 & 291.1 & 0.0 & 0.0  & 240.6 & 131.7 & 341.8 \\

94 & 0.0 &  0.0 &  0.0  & 7255.4 & 139.3 & 240.5 & 325.12 & 7960.32 \\

82, 89 & 118.4 & 2.2 & 8071.7 & 0.0 & 4.4  & 0.4 &  803.2 & 9000.3 \\

89, 94&  0.0 & 186.8 & 6.1  & 190.6 & 1.5 & 7365.8 & 488.2 & 8239 \\

82, 94 & 163.1& 0.0 &  9.5& 97.0 & 2662.2 & 0.09 & 331.5 & 3263.39 \\

AY & 62.5 &1.6 &164.4& 28.9 & 51.7 & 10.6 & 15923.7 & 18342.02\\ \hline

 \hline
NLTCS  &  8396  & 2959 
&  4464 
& 7572 
&  1511
& 3929  
&  6114 \\ 

\end{tabular}
\end{center}
\caption{Confusion Matrix for NLTCS}
\label{confusion}
\end{table}

\end{document}